\documentclass[11pt,twoside]{article}

\usepackage{edwin}
\usepackage{epsf}
\usepackage{psfig}
\usepackage{lscape}

\begin{document}
\title{E-prints and Journal Articles in Astronomy: a Productive Co-existence}   
\author{Edwin A. Henneken$^1$, Michael J. Kurtz$^1$, Simeon Warner$^2$, Paul Ginsparg$^2$, Guenther Eichhorn$^1$, Alberto Accomazzi$^1$, Carolyn S. Grant$^1$, Donna Thompson$^1$, Elizabeth Bohlen$^1$, Stephen S. Murray$^1$}
\affil{$^1$Harvard-Smithsonian Center for Astrophysics, 60 Garden Street, Cambridge, MA 02138}
\affil{$^2$Cornell Information Science, Ithaca, NY 14850, USA}

\begin{abstract} 

Are the e-prints (electronic preprints) from the arXiv repository being used instead of the journal articles? In this paper we show that the e-prints have not undermined the usage of journal papers in the astrophysics community. As soon as the journal article is published, the astronomical community prefers to read the journal article and the use of e-prints through the NASA Astrophysics Data System drops to zero. This suggests that the majority of astronomers have access to institutional subscriptions and that they choose to read the journal article when given the choice. Within the NASA Astrophysics Data System they are given this choice, because the e-print and the journal article are treated equally, since both are just one click away. In other words, the e-prints have not undermined journal use in the astrophysics community and thus currently do not pose a financial threat to the publishers. We present readership data for the arXiv category "astro-ph" and the 4 core journals in astronomy (Astrophysical Journal, Astronomical Journal, Monthly Notices of the Royal Astronomical Society and Astronomy \& Astrophysics). Furthermore, we show that the half-life (the point where the use of an article drops to half the use of a newly published article) for an e-print is shorter than for a journal paper.

The ADS is funded by NASA Grant NNG06GG68G. arXiv receives funding from NSF award \#0404553.
\end{abstract}

\section{Introduction}

The questions we address in this paper are twofold. The first question is a practical one: are people reading the e-prints from arXiv instead of the journal articles? In other words: are the journals safe? The second question concerns bibliometrics: Are e-prints read in a different way than journal articles? 

Besides providing easy access to technical literature, digital repositories like the NASA Astrophysics Data System (ADS;\citet{kurtz00}) and the arXiv e-print repository (\citeauthor{ginsparg94} \citeyear{ginsparg94} and \citeyear{ginsparg01}) also allow a detailed study of the readership of individual articles. This is the electronic equivalent of studying traditional library circulation records, listing who accessed what book when (\citet{kurtz05b}). We are interested in trends, however, not the usage of individual users.

The teams of the ADS and arXiv have been collaborating since 1997. In 1997 the ADS created an index for the astro-ph e-prints and made it available through the ADS abstract service with the same searching options as the published literature database. The abstracts of all arXiv categories were included in 2002. The incorporation of the arXiv abstracts resulted in the powerful alerting service {\it myADS-arXiv} (\citet{henneken06a}). Since mid-2004, arXiv usage data are incorporated in the ADS "also-read" statistics. The collaboration provides the unique opportunity to study differences between the ADS and arXiv readership. In other words: is there a difference between the "life" of a journal article and that of an e-print on arXiv? The "life" of an article can be characterized in various ways. The number of reads over time is one such way. Because of the stochastic nature of individual article reads, the "life" of an article only makes sense in an average sense.

The readership of the technical astronomical literature, available through the ADS, has already been studied extensively (\citet{kurtz05b},\citet{kurtz03}). 
In these studies article readership was measured by looking at how usage declines as a function of article age. For the astronomical literature (going back to the 1890s), a model was introduced describing 4 components ("modes") of readership. This model provides a good fit with the data. Each of the "modes" is characterized by a time scale, which is the "half-life" for that mode: the time where the use of an article drops to half the use of a newly published article. The different "modes" correspond with different types of usage. For example, people using the repository to look for historical litature form one group with specific usage characteristics. In general, one could say that articles move through different types of readership over time.  The model described in \citet{kurtz05b} introduces 4 distinct stages or "modes", where an article moves to "new" to "recent" to "interesting" to "historical".

Using the arXiv usage logs, we will investigate whether the readership of e-prints follows trends similar to the astronomical literature. Obviously, there will be differences because the articles of the Astrophysical Journal go back to the 1890s and the e-prints from arXiv go back to 1991. So, there is no "historical" component in the arXiv readership data. The most interesting time frame will be the first couple of months to a year after the publication of an e-print. The readership data from the arXiv logs will show whether e-prints are read in a different way than journal articles. Besides this bibliometric component, there is also an economic component: it has been said that the presence of e-prints is costing the publishers money. If astronomers read fewer journal articles because of the presence of e-prints, this might be true. At the same time, the opposite might be true as well. Maybe, thanks to the e-prints, journal articles have gained visibility and therefore an increased readership.

\section{Data}

\begin{figure}[!ht]
  \plotone{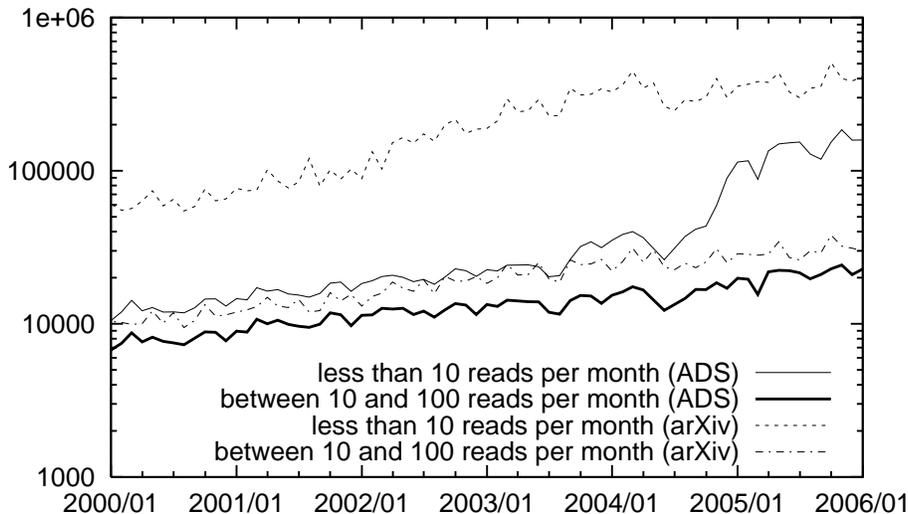}
  \caption{Classes of ADS and arXiv users. arXiv usage data available since June 2004}
  \label{fig:users}
\end{figure}

The source of our data consists of the Astrophysics Data System usage logs and usage data provided by the arXiv team (going back to the beginning of 2000). For the ADS, we log all types of access by our users. An access "type" is related to which type of information viewed for an article. We define "reads" as the access events by users, where multiple information retrievals per log period for one article by one user is regarded as a single "read". To rule out incidental use (e.g. by one-time users coming in via an external search engine, such as Google), we have taken the subset of users who query the database between 10 and 100 times per month. We took a similar slice of the arXiv usage data. We will refer to this set of users as the "set of typical users". The number of users in this set is around 20,000 per month the ADS and 30,000 per month for arXiv during the periods considered in this paper. This is illustrated in figure~\ref{fig:users}. Up to the middle of 2004, the number of ADS users doubled on a bi-yearly basis. Since Google started to index both the ADS and arXiv, the number of incidental users has increased dramatically. However, the number of typical users (10--100 reads per month) has continued to follow the same doubling pattern for both the ADS and arXiv. 

\section{Results}

\begin{figure}[!ht]
  \plotone{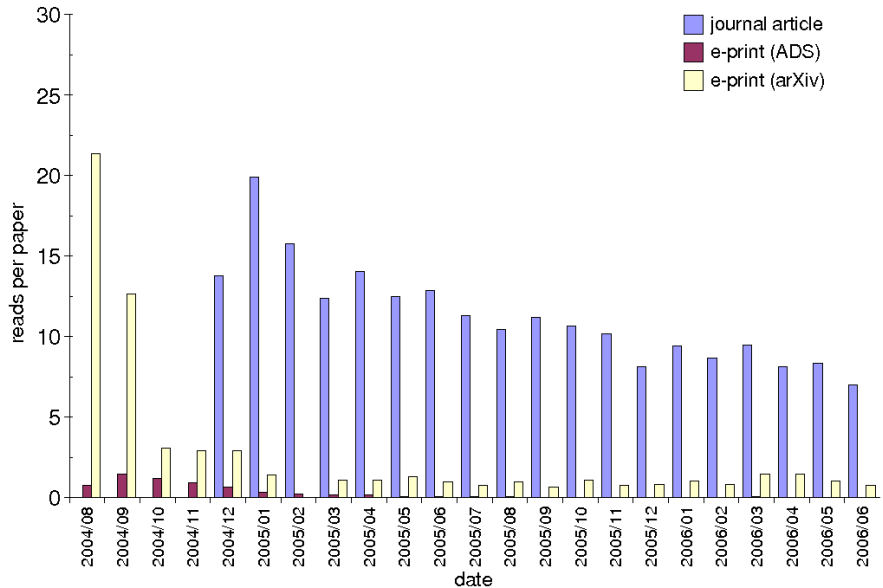}
  \caption{{\it C4} articles from December 2004, published 4 months after the arXiv e-print. Reads per paper from August 2004 through June 2006.}
  \label{fig:histo}
\end{figure}

We first take the set of all articles from the 4 core journals in astronomy (Astrophysical Journal, Astronomical Journal, Monthly Notices of the Royal Astronomical Society and Astronomy \& Astrophysics), from December 2004, for which the associated e-print was published in August 2004. This set consists of 118 papers. Figure~\ref{fig:histo} shows the reads per paper for the period of August 2004 through June 2006. From now on, we will refer to this set of the 4 core journals in astronomy as {\it C4}.

Two facts immediately follow from looking at figure~\ref{fig:histo}: (a) Reads of the arXiv e-print through the ADS very quickly drops to 0 after the publication of the journal article and (b) the half-life of the e-prints (the point where the use of an article drops to half the use of a newly published article) is a lot shorter than that of journal articles.

We have analyzed the same period, one year later, and the conclusion are the same. The only difference is that in that period, the reads of the e-print through the ADS immediately stop after the publication of the journal article, instead of gradually dissipating. 

The first fact immediately answers our first question: are the journals safe? The answer is yes. The ADS treats the e-prints and the journal articles equally, so the astronomical community is given the choice to read the e-print or the journal article. Figure~\ref{fig:histo} shows that the typical users prefer to read the journal article when this becomes available. This is an important observation. We believe this is because the journal article has been refereed and is accepted as the `official' version. This is good news for the publishers: e-prints have not undermined journal use in the astrophysics community and thus do not pose a threat to the journal readership. The fact that astronomers choose to read the journal article shows that they have access to institutional subscriptions, thanks to the librarians who make decisions about maintaining those subscriptions.

The second fact shows that the arXiv e-prints are heavily read in the month of their publication, quickly dropping off to a virtually constant readership of about 2 reads per paper after about 5 months. The reads per paper for the journal article only drop off gradually over this period. This behavior is the difference between browsing through a set of newly published e-prints and the intentional, large-scale reading of journal articles. 

\begin{figure}[!ht]
  \plotone{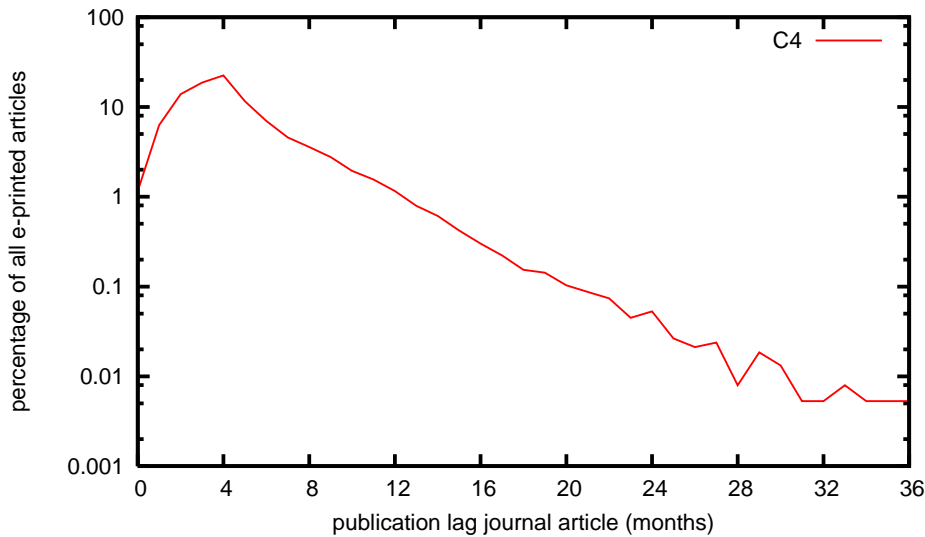}
  \caption{Number of months between e-print and journal article for all articles from the {\it C4} journals that also appeared as e-print on arXiv}
  \label{fig:lag}
\end{figure}

Our choice of 4 months lag between the publication dates of the e-print and the journal article is not coincidental. For the set of articles in {\it C4} which also appeared as e-prints on arXiv, the majority has a lag of 4 months (see figure~\ref{fig:lag}). 

\begin{figure}[!ht]
  \plotone{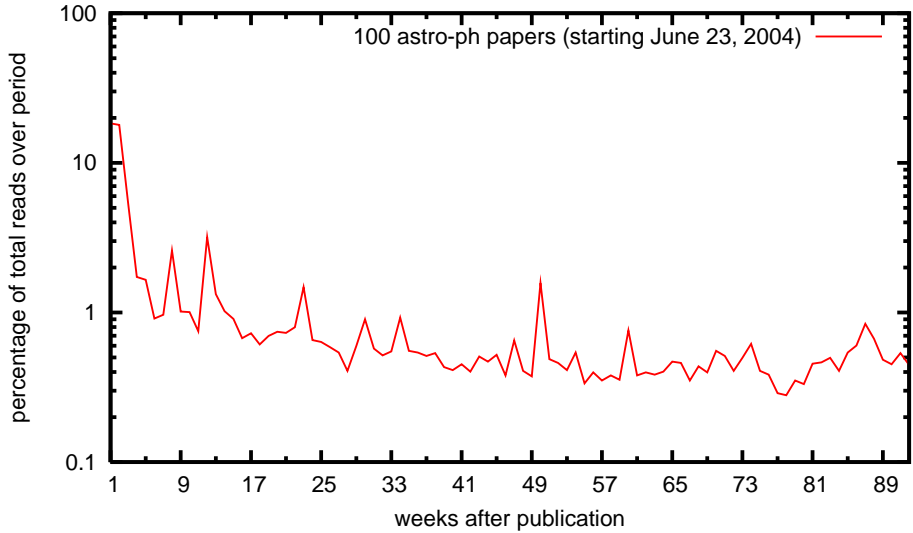}
  \caption{Weekly reads for 100 astro-ph papers from June 2004}
  \label{fig:weekly}
\end{figure}

Figure~\ref{fig:histo} has a resolution of one month. The data in this figure suggest that this is not enough resolution to get an accurate feeling of the readership half-life for e-prints. For this reason we decided to zoom in on the e-print readership by increasing the resolution to a week. We took a set of 100 astro-ph papers, all published within one week of June 2004 and monitored their reads up to June 2006. Figure~\ref{fig:weekly} shows this time series for the reads as percentage of the total number of reads over that period.

About 40\% of all reads took place within the first two weeks of the life of the e-print over that period.

\begin{figure}[!ht]
  \plotone{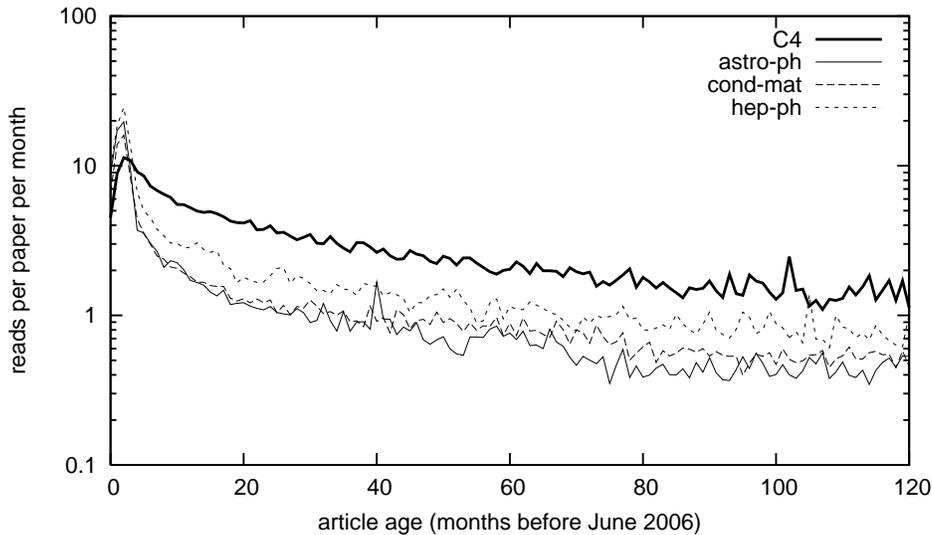}
  \caption{The number of reads per paper per month for three data sets}
  \label{fig:readprob}
\end{figure}

To illustrate the second fact, the dropoff of arXiv readership observed in figure~\ref{fig:histo}, we look at the data in another way. Now we will look at the set of all e-prints in "astro-ph" and all articles from the {\it C4} journals. In order to be able to make a fair comparison, we only look at the articles that appeared as e-prints in the "astro-ph" category. So, each element in one set is linked to one and only one element in the other set. This set contains 38,171 papers. Again for the typical users, we determine the reads in April through June of 2006, for the astro-ph papers and the articles in {\it C4}. For each publication month, we divide the number of reads by the number of papers published in that month. In this way, we will get the number of reads per paper as a function of paper age.
Figure~\ref{fig:readprob} shows the results for both sets. We have added the set of all papers published in the "cond-mat" and "hep-ph" categories as comparison.

The results in figure~\ref{fig:readprob} are a different way of illustrating the second fact we observed in figure~\ref{fig:histo}. The data for "astro-ph", "hep-ph" and "cond-mat" in figure~\ref{fig:histo} are purely based on the arXiv usage logs. The zero point in this figure corresponds with the publication month for articles in the data set considered. The e-prints are heavily read just after their publication, but after that their readership quickly drops off to a fairly constant rate. On the other hand, the readership of journal articles stays high after their publication. The fact that the lines for journal article reads and for e-print reads parallel each other after about 12 months means that, from then on, the usage patterns evolve similarly, albeit with a factor of about 5 less e-print reads. The big difference is in the early stages. The early peak in reads when e-prints appear (the "new" mode of \citet{kurtz05b}) is higher and much sharper than in the journal reads, which we attribute to "table of contents browsing". The similarity between the arXiv and ADS readership after about 12 months shows that a significant amount of people use arXiv as their search engine. Readership trends in physics fall outside the scope of this paper, but we do note the interesting fact that the reads for "hep-ph" lie significantly above those for "astro-ph" and "cond-mat". In this respect it is noteworthy that SPIRES, an important search engine for the high-energy physics community, offers mainly links to just the arXiv e-prints.

\section{Discussion}

The members of the astronomical community consitute the lion's share of the journal article readers. We have shown (figure~\ref{fig:histo}) that when given the choice, they prefer to read the article over the e-print. We believe this is because the journal article has been refereed and is accepted as the `official' version. This is also a very important observation with respect to the economic component. The majority of astronomers has access to the online journals through institutional subscriptions. In other words, the e-prints have not undermined journal use in the astrophysical community! Quite the contrary, the e-prints help journal articles to gain more visibility. Additional reasons for choosing the journal article are the cross-linking of references and datasets, the links to astronomical objects (SIMBAD) and the availability of high-resolution and low-resolution versions of plots and plates. In terms of the 4 modes, this means that the "interesting" and "historical" modes develop with smaller amplitudes in the readership of e-prints, because after the publication of the journal article, only a small subset of the typical users keeps reading the e-prints. The readership study in \citet{kurtz05a} has shown that the "new" mode has a half-life of about 16 days, the "current" mode has a half-life of about 1.7 years and the "interesting" mode has a half-life of about 10.7 years. The "historical" mode is a constant number of reads of about 1.5 reads per paper per month. For the e-prints, the "historical" mode seems settle at around 0.5 reads per paper per month. 

The e-prints and the journal articles co-exist and show vastly different usage characteristics. Their co-existence is productive, because the prior publication as e-print gives a journal article primacy. This primacy is one of the factors that results in higher citation rates (\citet{henneken06b} and references herein). 


\end{document}